\documentclass{revtex4}
\usepackage{graphicx}
\usepackage{amsmath}
\usepackage{amssymb}

\begin{document}
\title{Thick brane in 7D and 8D spacetimes}
\author{Vladimir Dzhunushaliev
\footnote{Senior Associate of the Abdus Salam ICTP}}
\email{dzhun@krsu.edu.kg} \affiliation{Dept. Phys. and Microel.
Engineer., Kyrgyz-Russian Slavic University, Bishkek, Kievskaya Str.
44, 720021, Kyrgyz Republic}

\author{Vladimir Folomeev}
\email{vfolomeev@mail.ru} \affiliation{Institute of Physics of NAS
KR, 265 a, Chui str., Bishkek, 720071,  Kyrgyz Republic}

\author{Kairat Myrzakulov}
\affiliation{Institute of Physics and
            Technology, 050032, Almaty, Kazakhstan}

\author{Ratbay Myrzakulov}
\email{cnlpmyra@satsun.sci.kz}
\affiliation{Institute of Physics and
            Technology, 050032, Almaty, Kazakhstan}

\begin{abstract}
We consider a thick brane model supported by two interacting scalar fields in
7D and 8D general relativity. Using the special type of a potential energy, we
obtain numerically the regular asymptotically flat vacuum solutions. A possibility
of obtaining the similar solutions for an arbitrary number of the extra spatial dimensions is estimated.
\end{abstract}

\maketitle

\section{Introduction}

The idea about the multidimensional Universe appeared in the 1920s for unification of
gravitational and electromagnetic interactions. Fast progress of the quantum theory
has overshadowed researches of multidimensional gravitational theories unifying
gravitation and other types of interactions. Just in the 1980s a great deal of interest has been
resumed within the framework of  superstring theory pretending to unify all of the fundamental
interactions in one theory. Perhaps the most important problem in the multidimensional theories is non-observability of extra dimensions. Firstly, it was proposed that extra dimensions are compactified and unobservable up to the Planck energy scales. Despite the numerous attempts for solving the problem of compactification of extra dimensions, one can think that the satisfactory solution of the problem was not found at the moment.

At the end of the 1990s, the idea that extra dimensions could be noncompactified (infinite),
but the matter is somehow concentrated on a 4D surface (brane), was proposed \cite{randall}
(see also the earlier works in this direction \cite{brane}). On the basis of this idea the new ways for
solution of some old problems of high-energy physics (the problem of mass hierarchy, stability of proton, etc.) were proposed.
At the moment, the models of 4D-branes embedded in 5D and 6D spaces \cite{gogberashvili2} and also
crossing 5D-branes in 6D \cite{Arkani-Hamed:1999hk} are under consideration.

It is supposed in brane models that various types of matter (bosons, fermions, gauge fields and so on)
are localized on a hypersurface (brane) embedded in an external multidimensional spacetime. However, in contrast
to the multidimensional Kaluza-Klein theory, extra dimensions could be macroscopic ones and, generally speaking,
non-compact ones. Another essential difference consists in different behavior of gravitational and matter fields:
if the former exists and propagates in the bulk, the matter fields are localized on a brane and they are
4D objects. At such an approach, however, the situation could be realized when a multidimensional
gravitational field could be also localized on a brane~\cite{Wang, Cast} (see also~\cite{Rubakov}), i.e.
effectively it becomes four dimensional despite the extra dimensions could be macroscopic ones. The
arising effective 4D gravitational constant is defined not only by physics on a brane but by influence of
extra dimensions also. These effects could be directly checked in experiments and observations both
on small and cosmological scales (see, e.g., Ref.~\cite{Rubakov}).

A special interest consists in consideration of various scalar fields within the framework
of the scenarios of the world on a brane. It is caused by that scalar fields are widely used
in particle physics, astrophysics and cosmology~\cite{Linde}. Within the framework of brane theory,
scalar fields were considered in different aspects in Refs.~\cite{Chamblin, Kanti, Sahni, Langlois, Kunze, Lee, Garousi}
(for a review, see~\cite{Brax}). This interest to the scalar fields is explained also by relative simplicity of
equations and solutions obtaining in models with using of them. It allows to make qualitative analysis of equations and
find sufficiently clear physical interpretation of results by analogy from other fields of physics.

One of the main problem at consideration of brane models is that in
almost all brane models one uses infinitely thin branes that, of
course, is absolutely unsatisfactory from the physical point of
view. The point is that obtaining of self-consistent solutions of
gravitational equations interacting with matter and describing a
thick brane is not a simple mathematical problem. At the moment
there exist the following thick-brane solutions: in
Ref.~\cite{Bronnikov:2005bg} such solutions are obtained as
monopole-like solutions. The set of gravitating scalar fields  with
non-trivial topological configuration far from the brane is
considered. It was shown that such solutions exist for co-dimension
of the brane more or equal to $2$. Unfortunately, for co-dimension
equal to 1 there is no topologically non-trivial configurations and,
consequently, there is no thick-brane solutions in this context.
Also in Ref~\cite{Delsate} the monopole-like solutions
with some potential of a scalar field with broken symmetry were considered. The regular or
periodic solutions for a global monopole with an arbitrary number of the extra spatial dimensions
$n$ and for a local monopole with $n=3$ were obtained.
In Ref.~\cite{Barbosa-Cendejas:2005kn} a thick-brane solution
in non-Riemannian  generalization of the 5D Kaluza-Klein theory was
obtained. In such an approach, affine connection, except Christoffel
symbols, has also the Weyl term which is described by a scalar
field. In Refs.~\cite{DeWolfe:1999cp}, \cite{Bazeia:2004dh} the
thick-branes were obtained at special choice of a potential of a
scalar field. It has allowed to integrate once the field equations
for the scalar field. The essential factor of obtaining of these
solutions consists in existence of the potential of the scalar field
unbounded below, i.e. $V(\phi) \rightarrow -\infty$ as $\phi
\rightarrow \pm \infty$.

In the light of aforesaid, it becomes interesting to consider various types of multidimensional brane models with
scalar fields for the purpose of finding out of possibility for existence of localized solutions with
finite energy for different number of extra dimensions. In this paper we show that two nonlinear gravitating scalar fields
could create a 4D brane in 7D and 8D spacetimes. From the physical point of view the situation is the following:
an interaction potential of these fields is constructed in such a way that it has two local and two global minima.
It means that there are two different vacuums. The multidimensional space is filled by these scalar fields which
are located in that vacuum in which the scalar fields are in the local minimum. There is a defect in the form of a 4D brane
on the background of this vacuum.

For 5D case such a model was considered in Ref.~\cite{Dzhun}, for 6D one - in Ref.~\cite{Dzhun1}. The obtained solutions indicate
a possibility of localization of scalar fields of special type on a brane.
\par
The search of thick brane solutions is very complicated mathematical
problem. Indeed, only three type of solutions are known till now:
(a) thick branes with a special choice of a potential of a scalar
field \cite{DeWolfe:1999cp}; (b) thick branes with topologically
non-trivial scalar fields \cite{Bronnikov:2005bg} and (c) thick
brane with two nonlinear scalar fields \cite{Dzhun}. Thus the goal
of our paper is to extend the solutions found in
Ref's~\cite{Dzhun}~\cite{Dzhun1} to higher dimensions. We do not
consider trapping of any matter on such branes because it is an
independent and complicated problem demanding a separate
consideration. We hope to investigate trapping of matter in our
future investigations.

\section{General equations}
In general case of $D=4+n$ dimensional gravity the action can be written as follows~\cite{Singl}:
\begin{equation}
\label{Daction}
S = \int d^Dx\sqrt {^Dg} \left[ -\frac{M^{n+2}}{2}R + L_m
\right]~,
\end{equation}
where $M$ is the fundamental mass scale and $n$ is a number of extra dimensions.
As a source of matter fields $L_m$ we chose two interacting scalar fields $\varphi, \chi$
with the Lagrangian:
\begin{equation}
\label{lagrangian}
    L_m =\frac{1}{2}\partial_A \varphi \partial^A
\varphi+\frac{1}{2}\partial_A \chi \partial^A
\chi-V(\varphi,\chi)~,
\end{equation}
where the potential energy $V(\varphi,\chi)$ is:
\begin{equation}
\label{pot2}
    V(\varphi,\chi)=\frac{\Lambda_1}{4}(\varphi^2-m_1^2)^2+
    \frac{\Lambda_2}{4}(\chi^2-m_2^2)^2+\varphi^2 \chi^2-V_0.
\end{equation}
(This potential energy was obtained in~\cite{Dzhunushaliev:2006di} at approximate modeling of a condensate of
gauge field in the SU(3) Yang-Mills theory.) Here and further capital Latin indices run over $A, B =0, 1, 2, 3, ..., D$ and
small Greek indices $\alpha, \beta =0, 1, 2, 3$ refer to four
dimensions; $\Lambda_1, \Lambda_2$ are the self-coupling constants, $m_1, m_2$
are the masses of the scalar fields $\varphi$ è $\chi$, respectively; $V_0$ is an arbitrary
normalization constant which will be chosen below from physical reasons.

Use of two fields ensures presence of two global minima of the potential \eqref{pot2} at
$\phi~=~0, \chi~=~\pm~m_2$ and two local minima at $\chi~=~0, \phi~=~\pm~m_1$ for
values of the parameters $\Lambda_1, \Lambda_2$ used in the
paper. The conditions for existence of the
local minima are: $\Lambda_1>0, m_1^2>\Lambda_2 m_2^2/2$, and for the
global minima: $\Lambda_2>0, m_2^2>\Lambda_1 m_1^2/2$. The presence of these minima
has allowed to find solutions localized on the brane for 5D and 6D cases in Refs.~\cite{Dzhun},~\cite{Dzhun1}
when the solutions have tended asymptotically to one of the local minima.

Variation of the action \eqref{Daction} with respect to the
$D$-dimensional metric tensor $g_{AB}$ led to Einstein's
equations:
\begin{equation}
\label{EinsteinEquation}
R^{A}_B - \frac{1}{2}\delta^{A}_B R = \frac{1}{M^{n+2}} T^{A}_B,
\end{equation}
where $R^{A}_B$ and $T^{A}_B$ are the $D$-dimensional Ricci and the energy-momentum
tensors respectively. The corresponding scalar field equations could be obtained from \eqref{Daction}
by its variation with respect to the field variables $\varphi,\chi$. Then these equations will be:
\begin{equation}
\label{FieldEquations}
\frac{1}{\sqrt{^D \! g}}\frac{\partial}{\partial x^A}\left[\sqrt{^D \! g}\,\, g^{AB}
\frac{\partial (\varphi,\chi)}{\partial x^B}\right]=-\frac{\partial V}{\partial (\varphi,\chi)}.
\end{equation}

\section{7D case}
In this case $n=3$ and the metric could be chosen in the form~\cite{Singl}:
 \begin{equation}
\label{metric}
ds^2=\phi^2(r) \eta_{\alpha \beta} dx^{\alpha} dx^{\beta}-\lambda(r)(dr^2+r^2 d\Omega^2_2),
\end{equation}
where $\eta_{\alpha \beta}$ is a flat 4D spacetime metric and the metric functions
 $\phi$ and $\lambda$ depend only on the extra coordinate $r$.
$d\Omega^2_2=d\theta^2+\sin^2\theta d\psi^2$ is an angular part of the metric depending only on 6-th and 7-th
coordinates.

Then, using \eqref{lagrangian},\eqref{EinsteinEquation} and \eqref{metric}, one can get the system of gravitational
equations in the following form:
\begin{eqnarray}
\label{Ein-7d-1}
3\left(2\frac{\phi^{\prime \prime}}{\phi}-\frac{\phi^\prime}{\phi}\frac{\lambda^\prime}{\lambda}\right)+
6\left(\frac{\phi^\prime}{\phi}\right)^2+
2\left\{3\frac{\phi^\prime}{\phi} \frac{(r^2 \lambda)^\prime}{r^2\lambda}+
\frac{(r^2 \lambda)^{\prime \prime}}{r^2\lambda}-\frac{1}{4}\left[\frac{(r^2 \lambda)^\prime}{r^2\lambda}\right]^2-
\frac{1}{2}\frac{\lambda^\prime}{\lambda}\frac{(r^2 \lambda)^\prime}{r^2\lambda}-\frac{1}{r^2}\right\} &=&
-\frac{2\lambda}{M^5}T^\alpha_\alpha, \\
\label{Ein-7d-2}
12\left(\frac{\phi^\prime}{\phi}\right)^2+2\left\{4\frac{\phi^\prime}{\phi} \frac{(r^2 \lambda)^\prime}{r^2\lambda}+
\frac{1}{4}\left[\frac{(r^2 \lambda)^\prime}{r^2\lambda}\right]^2-\frac{1}{r^2}\right\} &=&
-\frac{2\lambda}{M^5}T^r_r, \\
\label{Ein-7d-3}
4\left(2\frac{\phi^{\prime \prime}}{\phi}-\frac{\phi^\prime}{\phi}\frac{\lambda^\prime}{\lambda}\right)+
12\left(\frac{\phi^\prime}{\phi}\right)^2+4\frac{\phi^\prime}{\phi} \frac{(r^2 \lambda)^\prime}{r^2\lambda}+
\frac{(r^2 \lambda)^{\prime \prime}}{r^2\lambda}-
\frac{1}{2}\left[\frac{(r^2 \lambda)^\prime}{r^2\lambda}\right]^2-
\frac{1}{2}\frac{\lambda^\prime}{\lambda}\frac{(r^2 \lambda)^\prime}{r^2\lambda} &=&
-\frac{2\lambda}{M^5}T^\theta_\theta,
\end{eqnarray}
where a prime denotes differentiation with respect to $r$ and the corresponding components of the energy-momentum tensor are:
\begin{eqnarray}
\label{EMT-7d-1}
T^\alpha_\alpha&=&T^\theta_\theta=\frac{1}{2\lambda}\left(\varphi^{\prime 2}+\chi^{\prime 2}\right)+V(\varphi, \chi), \\
\label{EMT-7d-2}
T^r_r&=&-\frac{1}{2\lambda}\left(\varphi^{\prime 2}+\chi^{\prime 2}\right)+V(\varphi, \chi),
\end{eqnarray}
Multiplying the equation \eqref{Ein-7d-3} by $3/4$ and subtracting its from \eqref{Ein-7d-1}, and also
subtracting the equation \eqref{Ein-7d-2} from \eqref{Ein-7d-3}, we have:
\begin{eqnarray}
\label{Ein-n-7d-1}
\frac{\lambda^{\prime \prime}}{\lambda}-\frac{3}{5}\left(\frac{\lambda^\prime}{\lambda}\right)^2-
\frac{12}{5}\left(\frac{\phi^\prime}{\phi}\right)^2+
\frac{4}{5 r}\left(6\frac{\phi^\prime}{\phi}+
\frac{13}{4}\frac{\lambda^\prime}{\lambda}\right)+\frac{12}{5}\frac{\phi^\prime}{\phi}\frac{\lambda^\prime}{\lambda}&=&
-\frac{2}{5}\lambda\left[\frac{1}{2\lambda}\left(\varphi^{\prime 2}+\chi^{\prime 2}\right)+V(\varphi, \chi)\right], \\
\label{Ein-n-7d-2}
8\left(\frac{\phi^{\prime \prime}}{\phi}-\frac{\phi^\prime}{\phi}\frac{\lambda^\prime}{\lambda}-
\frac{\phi^\prime}{r \phi}\right)+\frac{\lambda^{\prime \prime}}{\lambda}-\frac{3}{2}\left(\frac{\lambda^\prime}{\lambda}\right)^2-
\frac{\lambda^\prime}{r \lambda}&=&-2\left(\varphi^{\prime 2}+\chi^{\prime 2}\right),
\end{eqnarray}
where we have used the following rescalings:
 $r \rightarrow r/M^{5/2}$, $\varphi \rightarrow M^{5/2} \varphi$,
$\chi \rightarrow M^{5/2} \chi$, $m_{1,2} \rightarrow M^{5/2} m_{1,2}$.
These equations are supplemented with two equations for the scalar fields which are obtained from
\eqref{FieldEquations}  with use of \eqref{metric}:
\begin{eqnarray}
\label{Field-Eq-1}
\varphi^{\prime \prime}+\left(\frac{2}{r}+4\frac{\phi^\prime}{\phi}+\frac{1}{2}\frac{\lambda^\prime}{\lambda}\right)\varphi^\prime
=\lambda \varphi\left[2\chi^2+\Lambda_1(\varphi^2-m_1^2)\right], \\
\label{Field-Eq-2}
\chi^{\prime \prime}+\left(\frac{2}{r}+4\frac{\phi^\prime}{\phi}+\frac{1}{2}\frac{\lambda^\prime}{\lambda}\right)\chi^\prime
=\lambda \chi\left[2\varphi^2+\Lambda_2(\chi^2-m_2^2)\right].
\end{eqnarray}
The set of equations \eqref{Ein-n-7d-1}-\eqref{Field-Eq-2} is a set of nonlinear equations with solutions
whose behavior depends essentially from values of the parameters $m_1, m_2, \Lambda_1, \Lambda_2$.
As it was shown in Refs.~\cite{Dzhun},~\cite{Dzhun1}, specifying some values of the self-coupling constants
$\Lambda_1, \Lambda_2$, a problem of search of {\it regular} solutions for systems similar to \eqref{Ein-n-7d-1}-\eqref{Field-Eq-2}
reduced to evaluation of eigenvalues of the parameters $m_1, m_2$. Only for these values of the parameters the
{\it regular} solutions with {\it finite} energy exist.

We will search for such regular solutions in a whole space. As it is not possible to find analytical solutions,
we search for a solution numerically. In order to make a solution regular everywhere, let us consider a solution
near the brane, i.e. at $r \approx 0$. Let us search for this solution as a series:
\begin{eqnarray}
    \phi(r) &\approx& \phi_0 + \phi_2 \frac{r^2}{2} ,
\label{sec2-120}\\
    \lambda(r) &\approx& \lambda_0 + \lambda_2 \frac{r^2}{2} ,
\label{sec2-130}\\
    \varphi(r) &\approx& \varphi_0 + \varphi_2 \frac{r^2}{2} ,
\label{sec2-140}\\
    \chi(r) &\approx& \chi_0 + \chi_2 \frac{r^2}{2} ,
\label{sec2-150}
\end{eqnarray}
where the subscript 0 denotes a value of the function at $r=0$ and
the subscript 2 denotes a second derivative of the corresponding function at $r=0$.
Substitution of these series into the equations \eqref{Ein-n-7d-1}-\eqref{Field-Eq-2} gives the following
values for $\phi_2, \lambda_2, \varphi_2, \chi_2$:
\begin{eqnarray}
    \lambda_2 &=& \frac{1}{15} \lambda_0^2 V(0), \quad \phi_2 = - \frac{2}{15} \phi_0 \lambda_0 V(0),
\label{sec2-160}\\
    \varphi_2 &=& \frac{1}{3} \lambda_0 \varphi_0 \left[
        2 \chi_0^2 + \Lambda_1 \left( \varphi_0^2 - m_1^2 \right)
    \right],  \quad
    \chi_2 = \frac{1}{3} \lambda_0 \chi_0 \left[
        2 \varphi_0^2 + \Lambda_2 \left( \chi_0^2 - m_2^2 \right)
    \right].
\label{sec2-170}
\end{eqnarray}

The obtained equations  \eqref{Ein-n-7d-1}-\eqref{Field-Eq-2} could not be solved analytically,
and so we solve them numerically. But also at a numerical analysis there are some difficulties.
The point is that the preliminary numerical analysis shows that a solution regular everywhere exists
not for all $m_{1,2}$ but for some special values of this parameters. This means that we have deal with a nonlinear
eigenvalue problem. The functions $\varphi(r), \chi(r)$ are eigenfunctions, and the parameters $m_{1,2}$ are
eigenvalues, and the condition $\Lambda_1 \neq \Lambda_2$ should be satisfied. In particular, we chose
$\Lambda_1 = 0.1$, $\Lambda_2 = 1.0$.

\subsection{Numerical solution}

In this section we describe a method of numerical solution of the equations \eqref{Ein-n-7d-1}-\eqref{Field-Eq-2} in details.
As it was mentioned above, the special feature of the set of equations is that determining of the functions $\varphi$ and $\chi$
is a nonlinear eigenvalue problem. Usually, numerical solution of one ordinary differential equation for
search of eigenfunctions and eigenvalues is carrying out by the shooting method.  The point of the method is that one,
using a method of step-by-step approximation, try to find with necessary accuracy some eigenvalue at which an eigenfunction
is regular. For example, in such a way one can find a discrete energy spectrum of a particle in a 1D potential well of
an arbitrary shape. Unfortunately, this method does not work in a case when one has a eigenvalues and eigenfunctions problem
with two variables. It takes place in our case: we have two eigenfunctions $\varphi, \chi$ and two eigenvalues $m_{1,2}$.

Therefore we will search for a numerical solution in the following way. Searching for eigenvalues of the equations
\eqref{Field-Eq-1}-\eqref{Field-Eq-2}, we solve on each step only whether the equation \eqref{Field-Eq-1} or
\eqref{Field-Eq-2}. Since there is also another function in this equation, its value takes from a previous step.
Having found (with some accuracy) some values of the functions $\varphi, \chi$ and eigenvalues $m_{1,2}$, one
should insert them into the Einstein equations \eqref{Ein-n-7d-1}-\eqref{Ein-n-7d-2} for determining of
the metric functions $\phi, \lambda$. Then these functions should be inserted in one of the field
equations \eqref{Field-Eq-1}-\eqref{Field-Eq-2} which is solving as a eigenfunctions problem for the functions
$\varphi, \chi$ and eigenvalues $m_{1,2}$ once again. These iterations repeat necessary number of times for obtaining
of a solution in the range $r \in [ \Delta, r_f ]$, where $r_f$ is some final point on the axes $r$.

Note the following features of the above method of a solution search of the differential equations
\eqref{Ein-n-7d-1}-\eqref{Field-Eq-2}:

\begin{itemize}
    \item One could not start a numerical solution from $r=0$ because there are terms like $y(r)/r$
    in the equations \eqref{Ein-n-7d-1}-\eqref{Field-Eq-2}, where $y(r)$ is one of the functions
    $\phi, \varphi, \chi, \lambda$. Therefore one should start a numerical solution from $r = \Delta \neq 0$ and
    use the following boundary conditions:
\begin{eqnarray}
    \phi(\Delta) &\approx& \phi_0 + \phi_2 \frac{\Delta^2}{2} ,
\label{sec3-10}\\
    \lambda(\Delta) &\approx& \lambda_0 + \lambda_2 \frac{\Delta^2}{2} ,
\label{sec3-20}\\
    \varphi(\Delta) &\approx& \varphi_0 + \varphi_2 \frac{\Delta^2}{2} ,
\label{sec3-30}\\
    \chi(\Delta) &\approx& \chi_0 + \chi_2 \frac{\Delta^2}{2} .
\label{sec3-40}
\end{eqnarray}
    \item One should not try to find a numerical solution for large $r_f$ because any numerical method has some inaccuracy.
    This inaccuracy defines (by some unknown way) the interval $[\Delta, r_f]$ in which an exact solution differs from
    numerical one insignificantly.
    \item An eigenvalues problem is very sensible to calculated eigenvalues. For example, let us consider the equation
\begin{equation}
    \frac{d^2 y(x)}{d x^2} = 2 y (x)  [y (x) ^2 - e] .
\label{sec3-50}
\end{equation}
    This equation, being considered as an eigenvalues problem with the eigenfunction $y(x)$ and eigenvalue $e$, has an exact
    solution $y = \tanh (x), e = 1$. If one solves this equation numerically with initial conditions which are followed from
    the exact solution
\begin{equation}
    y(0) = 0, \quad y'(0) = 1 ,
\label{sec3-70}
\end{equation}
then one can see that for $x \approx 6$ the numerical solution differs significantly from the exact one.
\end{itemize}
Thus we chose the following algorithm of numerical solution of the equations \eqref{Ein-n-7d-1}-\eqref{Field-Eq-2}:
\begin{enumerate}
    \item A choice of a zeroth approximation for the function $\varphi(r)$ or $\chi(r)$.
    \item Numerical solution of the equation \eqref{Field-Eq-1} or \eqref{Field-Eq-2}
    by the shooting method with evaluation of the functions $\varphi(r)$ or $\chi(r)$ and $m_1$ or $m_2$.
    \item Numerical solution of the equation  \eqref{Field-Eq-2} or \eqref{Field-Eq-1}
    by the shooting method with evaluation of the functions $\chi(r)$ or $\varphi(r)$ and $m_2$ or $m_1$.
    \item Reiteration of the steps 2 and 3 necessary number of times.
    \item Substitution of the functions $\varphi(r)$ and $\chi(r)$ into the Einstein equations \eqref{Ein-n-7d-1} and \eqref{Ein-n-7d-2}
    and their solution by a usual numerical method.
    \item Substitution of the functions $\phi(r)$ and $\lambda(r)$ into the equations \eqref{Field-Eq-1} or \eqref{Field-Eq-2}
    and reiteration of the steps 2-5 necessary number of times.
    \item Check of the obtained solutions by solution of the initial equations \eqref{Ein-n-7d-1}-\eqref{Field-Eq-2}
    with use of the calculated on the steps 2-3 parameters $m_{1,2}$.
\end{enumerate}
For these calculations we use the program package {\it Mathematica} for numerical solution of the differential equations.

\subsection{Results}
Using the procedure described in the previous section, let us find a self-consistent numerical
solution of the system \eqref{Ein-n-7d-1}-\eqref{Field-Eq-2}.
We chose the following boundary conditions at $r=0$:
\begin{alignat}{2}
\label{ini1}
\varphi_0&=\sqrt{3},& \qquad \varphi^\prime_0&=0, \nonumber \\
\chi_0&=\sqrt{0.6},& \qquad \chi^\prime_0&=0, \nonumber \\
\phi_0&=1.0,&  \qquad \phi^\prime_0&=0,\\
\lambda_0&=1.0,& \qquad \lambda^\prime_0&=0,\nonumber
\end{alignat}
and $\Lambda_1=0.1, \Lambda_2=1.0$. The arbitrary constant $V_0$ is chosen as $V_0=(\Lambda _2/4) m_2 ^4$ for
zeroing of the energy density as $r\rightarrow \infty$.

\begin{figure}[ht]
\begin{minipage}[t]{.49\linewidth}
  \begin{center}
  \includegraphics[width=9cm]{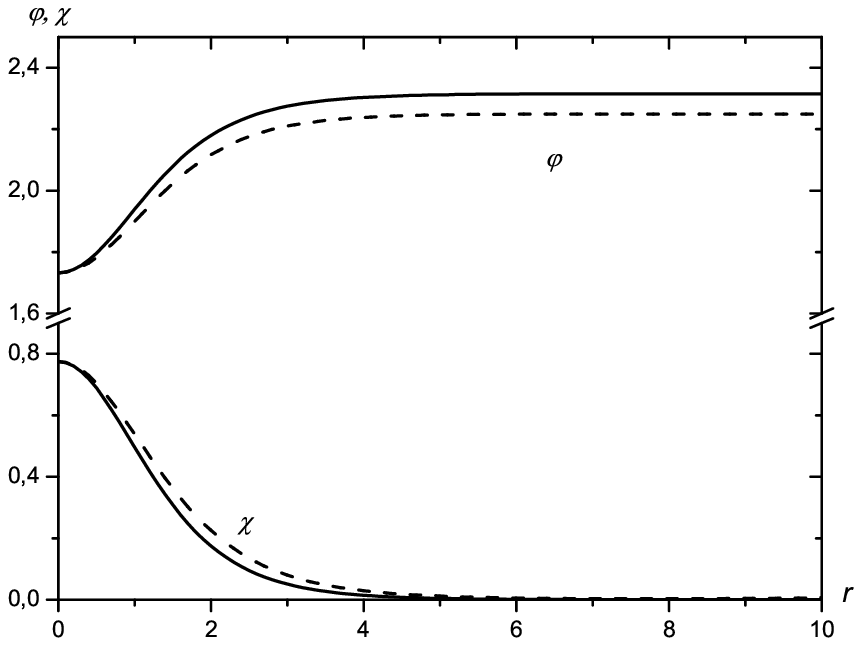}
  \caption{The scalar fields $\varphi(r), \chi(r)$ for the boundary conditions given in \eqref{ini1}.
  The solid lines correspond to the 7D case, the dashed lines - 8D case.}
    \label{phch1}
  \end{center}
\end{minipage}\hfill
\begin{minipage}[t]{.49\linewidth}
  \begin{center}
  \includegraphics[width=9cm]{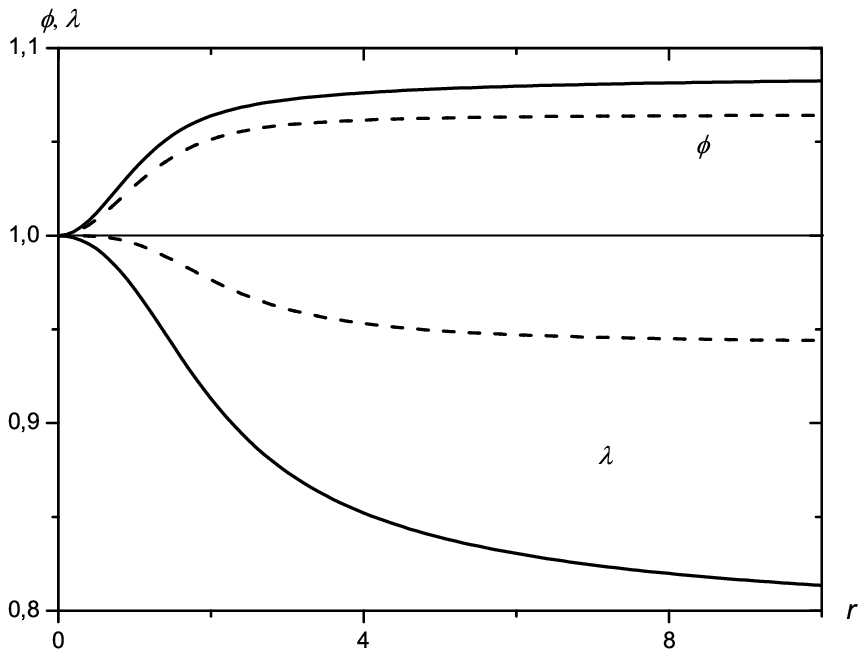}
  \caption{The metric functions $\phi(r), \lambda(r)$ for the boundary conditions given in \eqref{ini1}.
   The solid lines correspond to the 7D case, the dashed lines - 8D case.}
  \label{met1}
  \end{center}
\end{minipage}\hfill
\end{figure}

\begin{figure}[ht]
\begin{center}

  \includegraphics[width=9cm]{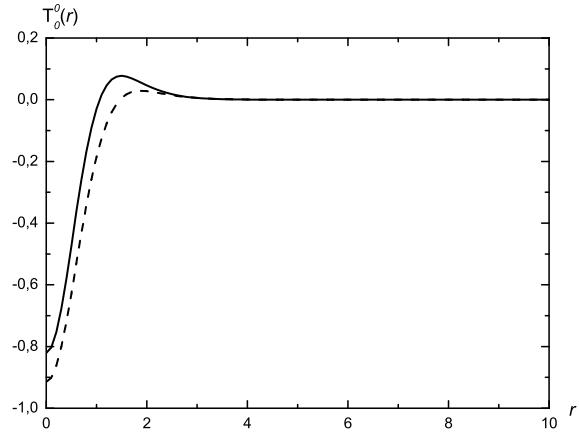}
 \caption{The energy density $T_0^0 (r)$ from \eqref{EMT-7d-1}.
  The solid line corresponds to the 7D case, the dashed line - 8D case.}
\label{energ1}
\end{center}
\end{figure}

In this case regular solutions exist at $m_1\approx 2.31505626$ and $m_2\approx 3.08288116$.
The obtained solutions are presented in Figs.~\ref{phch1}, \ref{met1}.
The corresponding plot for the energy density is shown in Fig.~\ref{energ1}.
As one can see from the last figure, the energy density tends asymptotically to zero as
$r\rightarrow \infty$, i.e. the scalar fields are trapped on the 4D-brane.

It is difficult to see an asymptotic behavior of the metric functions from Fig.~\ref{met1}.
For clarification of this question, let us find asymptotic solutions of the equations \eqref{Field-Eq-1}-\eqref{Field-Eq-2}
in the form:
\begin{equation}
\label{asymptotic}
\varphi=m_1-\delta \varphi, \quad \chi=\delta \chi,
\end{equation}
where $\delta \varphi, \delta \chi \ll 1$ as $r \rightarrow \infty$.
Such the asymptotic behavior of the fields corresponds to tending of the solutions to the local minimum at $\varphi~=~m_1, \chi~=~0$.
Then the right hand sides of the equations \eqref{Ein-n-7d-1}-\eqref{Ein-n-7d-2} tend to zero, i.e. one has the following
asymptotic equations for the metric functions:
\begin{eqnarray}
\label{Ein-as-7d-1}
\frac{\lambda^{\prime \prime}}{\lambda}-\frac{3}{5}\left(\frac{\lambda^\prime}{\lambda}\right)^2-
\frac{12}{5}\left(\frac{\phi^\prime}{\phi}\right)^2+\frac{4}{5 r}\left(6\frac{\phi^\prime}{\phi}+
\frac{13}{4}\frac{\lambda^\prime}{\lambda}\right)+\frac{12}{5}\frac{\phi^\prime}{\phi}\frac{\lambda^\prime}{\lambda}&=&
0, \\
\label{Ein-as-7d-2}
8\left(\frac{\phi^{\prime \prime}}{\phi}-\frac{\phi^\prime}{\phi}\frac{\lambda^\prime}{\lambda}-
\frac{\phi^\prime}{r \phi}\right)+\frac{\lambda^{\prime \prime}}{\lambda}-\frac{3}{2}\left(\frac{\lambda^\prime}{\lambda}\right)^2-
\frac{\lambda^\prime}{r \lambda}&=&0.
\end{eqnarray}
Let us find solutions of these equations in the form:
\begin{eqnarray}
\label{as-phi}
\phi&\approx& \phi_{\infty}-\frac{C_1}{r^\alpha}, \\
\label{as-lambda}
\lambda&\approx& \lambda_{\infty}+\frac{C_2}{r^\beta}.
\end{eqnarray}
(The subscript ``$\infty$'' indicates the asymptotic value of the variable, $C_1>0$, $C_2>0$ and $\alpha, \beta$ are
some arbitrary constants.)
One can check that in the case under consideration $\alpha$ and $\beta$ are equal each other. Then
we have from \eqref{Ein-as-7d-1}-\eqref{Ein-as-7d-2}:
\begin{eqnarray}
\label{as-phi_1}
\phi&\approx& \phi_{\infty}-\frac{C_1}{r}, \\
\label{as-lambda_1}
\lambda&\approx& \lambda_{\infty}+\frac{C_2}{r}.
\end{eqnarray}
Inserting these solutions in \eqref{Field-Eq-1}-\eqref{Field-Eq-2} and taking into account \eqref{asymptotic},
we have:
\begin{eqnarray}
\label{as-varphi}
\delta \varphi^{\prime \prime}+\frac{2}{r}\delta \varphi^\prime&=&2 \lambda_{\infty}\Lambda_1 m_1^2 \delta \varphi,\\
\label{as-chi}
\delta \chi^{\prime \prime}+\frac{2}{r}\delta \chi^\prime&=& \lambda_{\infty}(2 m_1^2-\Lambda_2 m_2^2) \delta \chi
\end{eqnarray}
with regular solutions
\begin{eqnarray}
\delta \varphi &\approx&  C_{\varphi} \frac{\exp{\left(- \sqrt{2 \lambda_{\infty} \Lambda_1 m_1^2} \,\, r \right)}}{r},
\label{sol1} \\
\delta \chi &\approx&  C_{\chi}\frac{\exp{\left(- \sqrt{ \lambda_{\infty} (2 m_1^2-\Lambda_2 m_2^2)} \,\,r\right)}}{r},
\label{sol2}
\end{eqnarray}
where $C_{\varphi}, C_{\chi}$ are integration constants.
Thus one can see from \eqref{asymptotic}, \eqref{as-phi}, \eqref{as-lambda}, \eqref{sol1} and \eqref{sol2} that the
asymptotic behavior corresponds to a 7D Minkowski spacetime with the scalar fields energy density equal to zero.

\section{8D case}
Now let us consider the 8D problem. For this case $n=4$ and we chose the metric as follows:
\begin{equation}
    ds^2 = \phi^2(r) \eta_{\mu \nu} dx^\mu dx^\nu -
    \lambda(r) \left\{
        dr^2 + r^2 \left[
            d \theta^2 + \sin^2 \theta \left( d \psi^2 + \sin^2 \psi d \xi^2 \right)
        \right]
    \right\},
\label{sec2-70}
\end{equation}
here $\mu ,\nu = 0,1,2,3$ refer to four dimensions; $r, \theta, \psi, \xi$ are extra coordinates;
$\theta, \psi, \xi$ are polar angles on a 3D sphere;
$\eta_{\mu \nu} = \left\{ +1, -1, -1, -1 \right\}$ is the 4D Minkowski metric.
Inserting the metric \eqref{sec2-70} in  \eqref{EinsteinEquation}-\eqref{FieldEquations}, one can
obtain the following equations:
\begin{eqnarray}
    4 \left(
        \frac{\phi''}{\phi} - \frac{\lambda' \phi}{\lambda \phi} -
        \frac{\phi'}{r \phi}
    \right) + \frac{\lambda''}{\lambda} -
    \frac{3}{2} \left( \frac{\lambda'}{\lambda} \right)^2 - \frac{\lambda'}{r \lambda}
    &=& -\left[ {\varphi'}^2 + {\chi'}^2 \right] ,
\label{sec2-80}\\
    \frac{\lambda''}{\lambda} - 2 \left( \frac{\phi'}{\phi} \right)^2 +
    \frac{4}{r} \left( \frac{\phi'}{\phi} + \frac{\lambda'}{\lambda} \right) +
    2 \frac{\phi' \lambda'}{\phi \lambda} -
    \frac{1}{4} \left( \frac{\lambda'}{\lambda} \right)^2
    &=& - \frac{1}{3} \lambda \left[
      \frac{1}{2 \lambda} \left( {\varphi'}^2 + {\chi'}^2 \right) + V
    \right] ,
\label{sec2-90}\\
    2 \frac{{\phi^{\prime}}^2}{\phi^2 } +
    2 \frac{\phi' \lambda'}{\phi \lambda} +
    4 \frac{\phi' }{r \phi} + \frac{3}{2} \frac{{\lambda'}^2}{\lambda^2} +
    4 \frac{\lambda' }{r \lambda} &=&
    -\frac{1}{3} \lambda
        \left[-\frac{1}{2 \lambda} \left( {\varphi'}^2 +
        {\chi'}^2 \right) + V
    \right] ~, \label{sec2-95} \\
    \varphi'' + \left(
        \frac{3}{r} + 4 \frac{\phi'}{\phi} + \frac{\lambda'}{\lambda}
    \right) \varphi &=& \lambda \varphi \left[
        2 \chi^2 + \Lambda_1 \left( \varphi^2 - m_1^2 \right)
    \right] ,
\label{sec2-100}\\
    \chi'' + \left(
        \frac{3}{r} + 4 \frac{\phi'}{\phi} + \frac{\lambda'}{\lambda}
    \right) \chi &=& \lambda \chi \left[
        2 \varphi^2 + \Lambda_2 \left( \chi^2 - m_2^2 \right)
    \right]
\label{sec2-110},
\end{eqnarray}
where we have used the rescalings:
$r \rightarrow r/M^{3}$, $\varphi \rightarrow M^{3} \varphi$,
$\chi \rightarrow M^{3} \chi$, $m_{1,2} \rightarrow M^{3} m_{1,2}$.

We will search for solutions of this system by analogy with the 7D case. Using the same boundary
conditions~\eqref{ini1}, one can obtain the results presented in Figs. \ref{phch1}-\ref{energ1}.

It is interesting to estimate an asymptotic behavior of the solutions as $r\rightarrow \infty$. For this purpose
let us search for a solution of the equations \eqref{sec2-100}-\eqref{sec2-110} in the form \eqref{asymptotic}.
Then the right hand sides of the equations \eqref{sec2-80}-\eqref{sec2-95} tend to zero and we will search
for solutions for the metric functions in the form \eqref{as-phi}-\eqref{as-lambda}. One can obtain from
\eqref{sec2-80}-\eqref{sec2-95} that $\alpha=\beta=2$, i.e. we have the following asymptotic behavior of
the metric functions:
\begin{eqnarray}
\label{as-phi_8d}
\phi&\approx& \phi_{\infty}-\frac{C_1}{r^2}, \\
\label{as-lambda_8d}
\lambda&\approx& \lambda_{\infty}+\frac{C_2}{r^2}.
\end{eqnarray}
Inserting these solutions in \eqref{sec2-100}-\eqref{sec2-110} and taking into account \eqref{asymptotic},
we will have:
\begin{eqnarray}
\label{as-varphi_8d}
\delta \varphi^{\prime \prime}+\frac{3}{r}\delta \varphi^\prime&=&2 \lambda_{\infty}\Lambda_1 m_1^2 \delta \varphi,\\
\label{as-chi_8d}
\delta \chi^{\prime \prime}+\frac{3}{r}\delta \chi^\prime&=& \lambda_{\infty}(2 m_1^2-\Lambda_2 m_2^2) \delta \chi
\end{eqnarray}
with regular solutions
\begin{eqnarray}
\delta \varphi &\approx&  C_{\varphi} \frac{\exp{\left(- \sqrt{2 \lambda_{\infty} \Lambda_1 m_1^2} \,\, r \right)}}{r^{3/2}},
\label{sol1_8d} \\
\delta \chi &\approx&  C_{\chi}\frac{\exp{\left(- \sqrt{ \lambda_{\infty} (2 m_1^2-\Lambda_2 m_2^2)} \,\,r\right)}}{r^{3/2}},
\label{sol2_8d}
\end{eqnarray}
where $C_{\varphi}, C_{\chi}$ are integration constants.
Thus one can see from \eqref{asymptotic}, \eqref{as-phi_8d}, \eqref{as-lambda_8d}, \eqref{sol1_8d} and \eqref{sol2_8d} that,
such as in the 7D case, the asymptotic behavior corresponds to a 8D Minkowski spacetime with the zero scalar fields energy density.

\section{Arbitrary number of the extra spatial dimensions}
\label{gen_n}
Obtaining of the numerical regular solutions for the 7D and 8D cases
allows us to hope that similar solutions could exist and for an arbitrary number of the $n$ extra spatial dimensions.
Unfortunately, one cannot obtain numerical solutions for the arbitrary $n$ because it is not possible to
eliminate $n$ from the equations. However, we can estimate a possibility
of obtaining of such solutions. Let us use for this the Einstein equations obtained in Ref.~\cite{Singl} for
the generalized $D$-dimensional metric
\begin{equation}
\label{metric_n}
ds^2= \phi ^2(r) \eta_{\alpha \beta }(x^\nu)dx^\alpha dx^\beta -
\lambda (r) (dr^2 +  r^2 d \Omega ^2 _{n-1}) ~,
\end{equation}
where $d \Omega ^2 _{n-1}$ is the solid angle for the $(n-1)$ sphere.
It is convenient to rewrite the Einstein equations from~\cite{Singl} in the form:
\begin{eqnarray}
\label{Einstein-na}
    3 \left( 2\frac{\phi ^{\prime \prime}}{\phi} -
    \frac{\phi ^{\prime}}{\phi} \frac{\lambda ^{\prime}}{\lambda }
    \right) + 6 \frac{(\phi ^{\prime})^2}{\phi ^2} + &&
\nonumber \\
    (n-1) \left[
    3\frac{\phi ^{\prime}}{\phi}\left(\frac{\lambda^\prime}{\lambda}+\frac{2}{r}\right)+
    \frac{\lambda^{\prime\prime}}{\lambda}-
    \frac{1}{2}\frac{\lambda ^{\prime}}{\lambda}\left(\frac{\lambda^\prime}{\lambda}-
    \frac{6}{r}\right)+ \frac{n-4}{4}\frac{\lambda^\prime}{\lambda}
    \left(\frac{\lambda^\prime}{\lambda}+
    \frac{4}{r}\right)  \right]
    &=&-\frac{2\lambda}{M^{n+2}}T^\alpha_\alpha  ~,  \\
\label{Einstein-nb}
    12 \frac{(\phi ^{\prime})^2}{(\phi)^2 }+
    (n-1) \left[ 4 \frac{\phi ^{\prime}}{\phi}\left(\frac{\lambda^\prime}{\lambda}+
    \frac{2}{r}\right)+
    \frac{n-2}{4}\frac{\lambda^\prime}{\lambda} \left(\frac{\lambda^\prime}{\lambda}+
    \frac{4}{r}\right) \right] &=&
    -\frac{2\lambda}{M^{n+2}}T^r_r ~, \\
\label{Einstein-nc}
    4 \left( 2\frac{\phi ^{\prime \prime}}{\phi} -
    \frac{\phi ^{\prime}}{\phi} \frac{\lambda ^{\prime}}{\lambda }
    \right) + 12 \frac{(\phi ^{\prime})^2}{\phi ^2} + &&
\nonumber \\
    (n-2)\left[ 4\frac{\phi ^{\prime}}{\phi}\left(\frac{\lambda^\prime}{\lambda}+
    \frac{2}{r}\right)+
    \frac{\lambda^{\prime\prime}}{\lambda}-
    \frac{1}{2}\frac{\lambda^{\prime}}{\lambda}
    \left(\frac{\lambda^\prime}{\lambda}-\frac{6}{r}\right)+
    \frac{n-5}{4}\frac{\lambda^\prime}{\lambda}
    \left(\frac{\lambda^\prime}{\lambda}+\frac{4}{r}\right) \right]
    & = & -\frac{2\lambda}{M^{n+2}}T^\theta_\theta~,
\end{eqnarray}
and the scalar field equations are:
\begin{eqnarray}
  \varphi'' + \left(
        \frac{n-5}{r} + 4 \frac{\phi'}{\phi} + \frac{\lambda'}{\lambda}
    \right) \varphi &=& \lambda \varphi \left[
        2 \chi^2 + \Lambda_1 \left( \varphi^2 - m_1^2 \right)
    \right] ,\\
    \chi'' + \left(
        \frac{n-5}{r} + 4 \frac{\phi'}{\phi} + \frac{\lambda'}{\lambda}
    \right) \chi &=& \lambda \chi \left[
        2 \varphi^2 + \Lambda_2 \left( \chi^2 - m_2^2 \right)
    \right].
\end{eqnarray}
As one can see from the equations~\eqref{Einstein-na}-\eqref{Einstein-nc}, the left hand sides could be regular at $r=0$ if the boundary conditions
$\lambda^\prime(0)=0, \phi^\prime(0)=0$ are hold true. Just these conditions for the metric functions
were used by us at consideration of the 5D~\cite{Dzhun}  and 6D~\cite{Dzhun1} cases and also
the 7D and 8D problems in this paper. Apparently, the situation remains the same and for
a case of an arbitrary number of the $n$ extra spatial dimensions.

On the other hand, consideration of an asymptotic behavior of the metric functions $\phi, \lambda$ for the
7D and 8D cases shows that their behavior could be described for the arbitrary $n> 6$
in the following form:
\begin{eqnarray}
\phi&\approx& \phi_{\infty}-\frac{C_1}{r^{n-6}}, \\
\lambda&\approx& \lambda_{\infty}+\frac{C_2}{r^{n-6}}.
\end{eqnarray}
Such a behavior corresponds to fast transition of the solutions to a Minkowski spacetime. At the same time,
an asymptotic behavior of the scalar fields could be described as follows:
\begin{eqnarray}
\varphi &\approx&  m_1-C_{\varphi} \frac{\exp{\left(- \sqrt{2 \lambda_{\infty} \Lambda_1 m_1^2} \,\, r \right)}}{r^{(n-5)/2}},\\
\chi &\approx&  C_{\chi}\frac{\exp{\left(- \sqrt{ \lambda_{\infty} (2 m_1^2-\Lambda_2 m_2^2)} \,\,r\right)}}{r^{(n-5)/2}}.
\end{eqnarray}

So we have the regular solutions for the arbitrary $n$ both in neighborhood of zero and as $r\rightarrow \infty$.
Note that it is possible to expand the solutions near zero in a Taylor series (see \eqref{sec2-120}-\eqref{sec2-150})
and obtain solutions with any necessary accuracy.
One might expect that these solutions could be joined smoothly and they would be regular in the entire
range $0 \leq r < \infty$.

\section{Conclusions}
We have shown that two gravitating nonlinearly interacting scalar fields could form a 4-dimensional thick brane configuration in
7D and 8D spacetimes. Consideration of the problem reduced to investigation of the system of ordinary
differential equations. It was shown that the regular solutions with the finite energy density exist only for
some values of the masses of the scalar fields $m_1, m_2$ at some self-coupling constants
$\Lambda_1, \Lambda_2$ given beforehand. I.e. the problem of a search of {\it eigenvalues} of the parameters $m_1, m_2$ for the
system of nonlinear equations \eqref{EinsteinEquation}-\eqref{FieldEquations} was solved. There are regular solutions
with an asymptotically flat metric $\phi(\infty)~=~\phi_\infty, \lambda(\infty)~=~\lambda_\infty$ and with
the zero energy density of the matter fields at $m_1\approx 2.31505626$ and $m_2\approx 3.08288116$ (for the 7D case)
and  $m_1\approx 2.25005$ and $m_2\approx 3.115$ (for the 8D case).
In both cases the solutions start with the boundary conditions \eqref{ini1} with subsequent transition to the
local minimum of the potential \eqref{pot2}.

A numerical analysis shows that the solutions exist since the potential $V(\varphi, \chi)$ has the local minimum.
Our attempts to find a solution tending to the global minimum as $r \rightarrow \infty$ were unsuccessful.
The possible reason is the Derrick's theorem \cite{derrick} which forbids an existence of static regular solutions
with finite energy for scalar fields in dimensions more then 2. At numerical solution (on the step 4) we obtain a solution
which is regular both in a whole 7D and 8D spacetimes. At first sight this solution seems to be forbidden by
the Derrick's theorem but more careful analysis shows that the obtained solution avoids the conditions of the given theorem
because the scalar fields tend asymptotically to the local minimum but not to the global one.

Let us discuss a question about stability of the obtained solutions. If one consider these fields as classical ones, then
it is possible to test the stability by the following way: (a) first of all, one can investigate small perturbations by a standard
method; (b) then one should investigate large perturbations; (c) following which it is necessary to consider a behavior of
this solution after quantizing of the scalar fields. In the last case the quantum fields could tunnel from a region of the local
minimum (false vacuum) to a region of the global minimum (true vacuum). But in this case, apparently, the solution should
decay into some waves because the existence of a static solution with scalar fields tending asymptotically to a global
minimum is forbidden by the Derrick's theorem.

But situation could turn out to be more interesting: the point is that in Ref.~\cite{Dzhunushaliev:2006di} there are some
arguments in favour that the scalar fields considered in this paper are quantum non-perturbative condensate of SU(3) gauge field.
Briefly these arguments consist in the following: components of SU(3) gauge field could be divided by some natural way into two
parts. The first group contains those components which belong to a subgroup $SU(2) \in SU(3)$. The remaining components belong
to a factor space $SU(3)/SU(2)$. Non-perturbative quantization is carrying out by the following approximate way: it is supposed that
2-point Green functions could be expressed via the scalar fields. The first field $\varphi$ describes 2-point Green functions
for SU(2) components of a gauge potential, and the second scalar field $\chi$ describes 2-point Green functions
for $SU(3)/SU(2)$ components of a gauge potential. It is supposed further that 4-point Green functions could be obtained as some
bilinear combination of 2-point Green functions. Consequently, the Lagrangian of SU(3) gauge field takes the form \eqref{lagrangian}.

At such an interpretation, a question about stability of the obtained thick brane solution becomes rather nontrivial problem.
In this case the obtained solution describes some defect in a spacetime filled by a condensate of a gauge field.
A question about stability demands a non-perturbative quantized consideration of SU(3) gauge field with Green functions
depending on time. Difficulty consists that the suggested in Ref.~\cite{Dzhunushaliev:2006di} approximate method of
description of Green functions could be used only in a stationary case.

Finally, in section \ref{gen_n} we have discussed a possibility of obtaining of thick brane solutions with an arbitrary
number of the $n$ extra spatial dimensions. It was shown that there are regular solutions for regions near zero and as
$r\rightarrow \infty$ for the arbitrary $n$. It allows to hope for existence of solutions joining these regions smoothly.

The whole set of results obtained in works \cite{Dzhun} for 5D and \cite{Dzhun1} for 6D cases, and also in this paper,
allows us to speak about possibility in principle of localization of the scalar fields with the potential \eqref{pot2}
on a brane in any dimensions. One has every reason to suppose that existence of similar regular solutions is possible for models with
greater number of extra dimensions. In future we suppose to investigate models with arbitrary number of extra dimensions
for clarification of this question.

\end{document}